\documentclass[prb,preprint,showpacs,preprintnumbers,amsmath,amssymb]{revtex4}
\usepackage{graphicx}
\usepackage[citecolor=blue,colorlinks=true,linkcolor=blue]{hyperref}
\usepackage{dcolumn}

\begin{document}

\title{Island Size Selectivity  and island-shape analysis during 2D Island Coarsening of Ag/Ag (111) Surface }

\author{Giridhar Nandipati}
\email{giridhar.nandipati@ucf.edu	}
\author{Abdelkader Kara}
\email{abdelkader.kara@ucf.edu}
\author{Syed Islamuddin Shah}
\email{islamuddin@knights.ucf.edu}
\author{Talat S. Rahman}
\email{talat.rahman@ucf.edu}
\affiliation{Department of Physics, University of Central Florida,  Orlando, FL  32816}

\date{\today}

\begin{abstract}
In our earlier study of Ag island coarsening on Ag(111) surface using  kinetic Monte Carlo (KMC) simulations we found that during early stages coarsening proceeds as a sequence of selected island sizes resulting in peaks and valleys in the island-size distribution. \cite{iss_short}, and that this selectivity is independent of initial conditions and dictated instead by the relative energetics of edge-atom diffusion and detachment/attachment processes and by the large activation barrier for kink detachment.
In this paper we present a detailed analysis of the shapes of various island sizes observed during these KMC simulations and show that selectivity is due to the formation of kinetically stable island shapes which survive longer than non-selected sizes, which decay into nearby selected sizes. The stable shapes have a closed-shell structure -  one in which every atom on the periphery having at least three nearest neighbors.
Our KMC simulations were carried out using a very large database of processes identified by each atom's unique local environment, the activation barriers of which were calculated using semi-empirical interaction potentials based on the embedded-atom method. 
\end{abstract}
\maketitle
\section{Introduction}


The phenomenon of coarsening plays an important role in a wide variety of processes in many branches of the physical sciences. Of particular interest is coarsening of two- or three-dimensional islands on various surfaces. Given its technological importance, coarsening has been the subject of a great deal of experimental and theoretical investigation. \cite{Zinke, pai1,Khare, Meakin, Soler, sholl, Shollprl,Kandel, Zinke2, pslkmc, pkmc, OstwaldAg} Coarsening of islands at late stages is dominated by Ostwald ripening (OR),\cite{ostwald, voorhees, LS} driven by lowering of excess surface free energy associated with island edges. The result is that islands larger than a critical size grow at the expense of smaller islands. 
Since these islands are assumed to be immobile, coarsening is considered to be mediated by diffusion of atoms between islands resulting in asymptotic self-similar growth, with the characteristic linear dimension  $L$ increasing with time in accordance with $L \sim t^{1/3}$ (the Lifshitz-Slyozov law)\cite{LS,Wagner}.
In recent years the development of fast-scanning tunneling microscopes (STM) has triggered the investigation of changes in surface morphologies that were deliberately created far from equilibrium \cite{peale,theis1,theis2,morgenstern1,Wen,hannon,tanaka,giesen,morgenstern2} as well as	of thermal	fluctuation around equilibrium-shaped structures \cite{poensgen,kuipers,morgenstern3,pai1,pai2} in high-temporal resolution.
Room temperature studies using scanning tunneling microscopy (STM) confirm that Ag/Ag(111) coarsening at late stages is dominated by Ostwald ripening..\cite{morgenstern1,morgenstern2} 
In contrast to all previous studies which concentrate on late-stages when the islands are very large and coarsening is dominated by exchange of atoms between islands, here we study initial stages of coarsening, that is when the islands are small.

Recently, by the way of kinetic Monte Carlo simulations using a large database of processes, we have shown that during early stages of Ag(111) island coarsening i.e, when islands are smaller in size, coarsening proceeds as a sequence of selected island size.\cite{iss_short} This results in peaks at selected island sizes and valleys at non-selected island sizes in the island-size distribution (ISD). 
The island selectivity was found to be independent of choice of initial ISD, of the initial shape of islands and of surface temperature though, the {\it strength} of island-selectivity does depend on temperature. We found that the island of selected sizes do not have atoms diffusing along their edges.  The fact that all atoms thus have at least three nearest neighbors, makes them kinetically stable islands and explains the peaks in the ISD at these island sizes. In this article, we extended out investigation of  island size selectivity to detailed analysis of {\it shapes} of islands observed during initial stages of Ag island coarsening on Ag(111) surface, and discuss reasons why islands of certain sizes  do not form kinetically stable shapes even when one potentially exists. In contrast to a previous static study \cite{Ag111stuct} of Ag clusters on Ag(111),  we show that these kinetically stable shapes does not necessarily have to be low-energy shapes.

The organization of this paper is as follows. In Sec.~\ref{KMC} we briefly describe self-learning kinetic Monte Carlo and the database of processes that we used in our simulations. In Sec.~\ref{results} we present results of our simulations of the initial stages of Ag/Ag(111) island coarsening and our understanding of reasons for island-size selectivity. Specifically, we discuss how island-size selectivity is governed by the way relative energy barriers for various detachment processes and  determine the shape of islands of various sizes formed during coarsening results in island size selectivity. Finally, in Sec.~\ref{conclusions} we present our conclusions.

\section{Simulations}\label{KMC}

Kinetic Monte Carlo (KMC) is an extremely efficient method \cite{bkl,gilmer,voter1,maksym,kristen, Blue} for carrying out a wide variety of dynamical simulations of non-equilibrium processes when the relevant activated atomic-scale processes are known {\it a priori}. Accordingly, KMC simulations have been successfully used to model a variety of dynamical processes ranging from catalysis to thin-film growth over experimentally relevant length and time scales.

In our simulations we made use of  a very large database of processes obtained from previous self-learning KMC (SLKMC) \cite{slkmc1,slkmc2,offkmc} simulations of small and large Ag-island diffusion on an Ag(111) surface carried out at temperatures of $300$ and $500$ K. This database has wide variety of single, multi- and concerted atom processes. All diffusion processes in this database move atoms from one fcc site to another. 
 In an earlier study,\cite{pslkmc} we used this database to do long timescale KMC simulations (on the order of a few hundred seconds) of Ag island coarsening on Ag(111) surface at room temperature.  
 We used the embedded-atom method (EAM) as developed by Foiles {\it et al} \cite{Foiles} to build interaction potentials.
Since rates are not expected to be strongly affected in the low temperature regime explored, we introduced a  simplification by assuming a `normal' value for all diffusion prefactors, although we are aware that multi-atom processes may be characterized by high prefactors \cite{handan,pref1,pref2}. Rates are not expected to be strongly affected in the explored low/moderate temperature regime. In our simulation we assumed a prefactor of $10^{12}~$s$^{-1}$. 
Similar to our previous study\cite{pslkmc} we used relatively large system size of $1024 \times 1024$ fcc lattice units with periodic boundary conditions  in order to avoid finite size effects  and to get good statistics our results were averaged out $10$ runs.
More details about database acquisition, types of processes with their respective activation energy barriers, recipes for speeding KMC simulations and other additional details about the simulation can be found in Ref.\onlinecite{pslkmc}.

\section{Results}\label{results}
\subsection{Initial Configuration}

\begin{figure}[ht]
\center{\includegraphics [width=6.5cm]{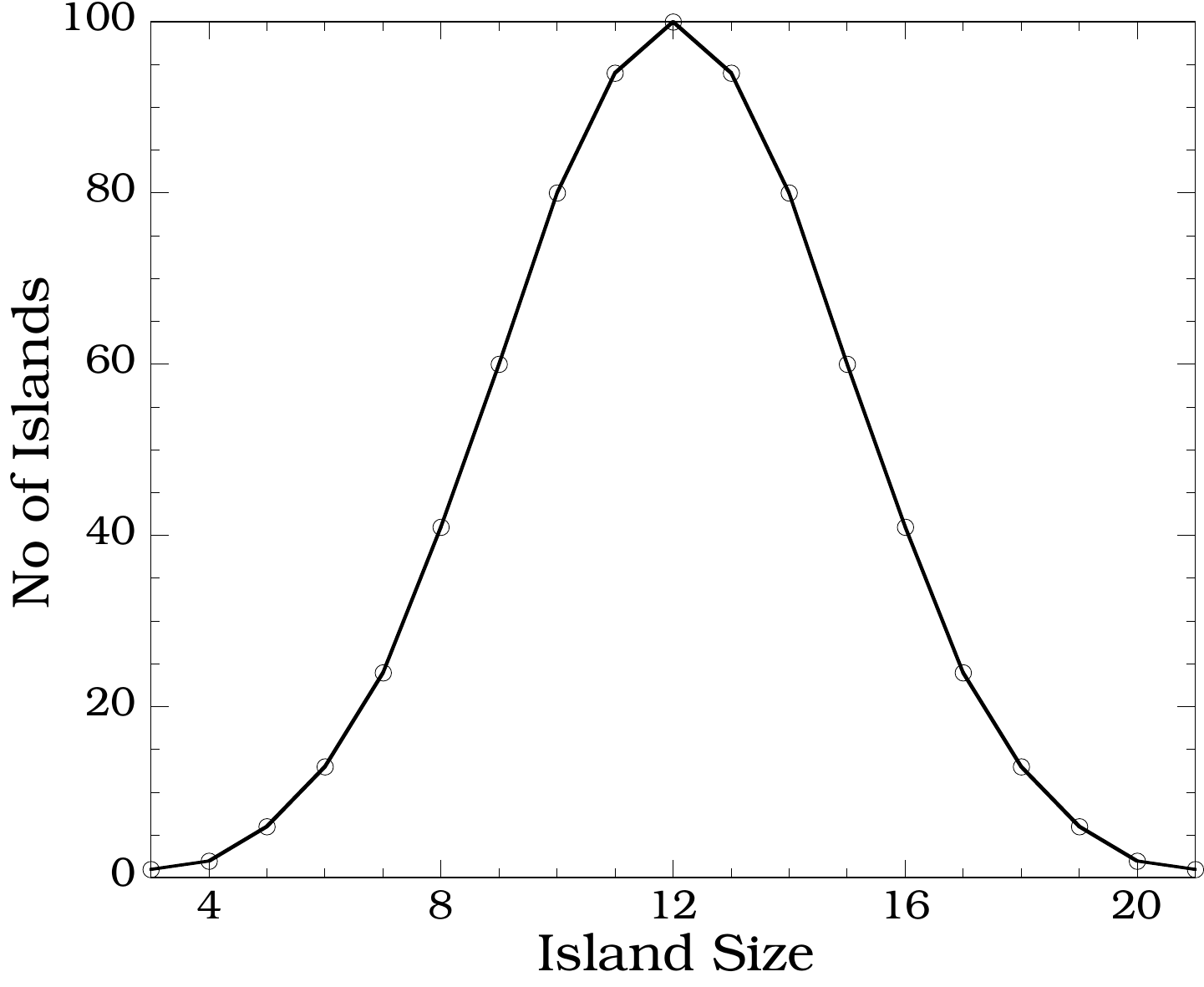} 
\caption{\label{gaussian}{Gaussian initial island size distribution with peak of 100 island at size $12$ and width of 3.}}}
\end{figure}

We created  the initial  distribution of islands for our coarsening simulation  by first dividing the  empty  lattice  into boxes of equal size, then  randomly  selecting an island size and a box, and placing an island of that size at the center  of that box, so as to prevent any overlap of islands.  In our case we divided the lattice into $1024$ boxes of size $32 \times 32$.  The total  number of islands and  the number  of islands of a particular size in the distribution depends on the  type of initial  ISD chosen.  We ran  our simulations  using both  Gaussian  and delta  initial  ISDs.  In the delta initial distributions, we set all $742$ islands at a given size (repeating the simulation for islands of all sizes between $10$ and $30$).  For a Gaussian  distribution the total number of islands depends on the number  of islands (a) of average  size Ñthat is, the number of islands at  the peak  of the  distribution ($\mu$)  -- and the width  of the distribution ($\sigma$).  All island sizes between $\mu \pm \sigma \sqrt{2~\ln(a)}$ are present in the distribution so that  the distribution is uniform around  the average island size. Fig.~\ref{gaussian} shows an example of a Gaussian  initial  island-size distribution, this one with peak of $100$ islands at  size $12$ and width  of $3$. 

For each initial Guassian ISD we arbitrarily set the peak at $100$ islands at the average size and set the width  at 3.  The result in each case for an initial  Gaussian  ISD is  a total  of $742$ islands , the set of initial distributions differing from each other in the number of atoms set as the average size.That is:  the total  number of islands in the initial  ISD is kept  constant $(742)$ for all  simulations  by keeping the peak island count and the width of the  Gaussian  distribution constant regardless of the  average island size.    For simplicity the shapes of islands in the initial  ISD were chosen arbitrarily, and islands  of same size were assigned the same shape.  The simulation was repeated for different shapes for a given island size.  For the results  presented  here, most of the initial island shapes were either compact  or close to compact.  To insure that our results were not a reflection of this choice of initial shapes, we ran another set of simulations in which the assigned shapes were fractal.  The results between the two sets of simulations were indistinguishable.

\subsection{Island-Size Selectivity}

 \begin{figure}[h]
 \includegraphics [width=8.5cm]{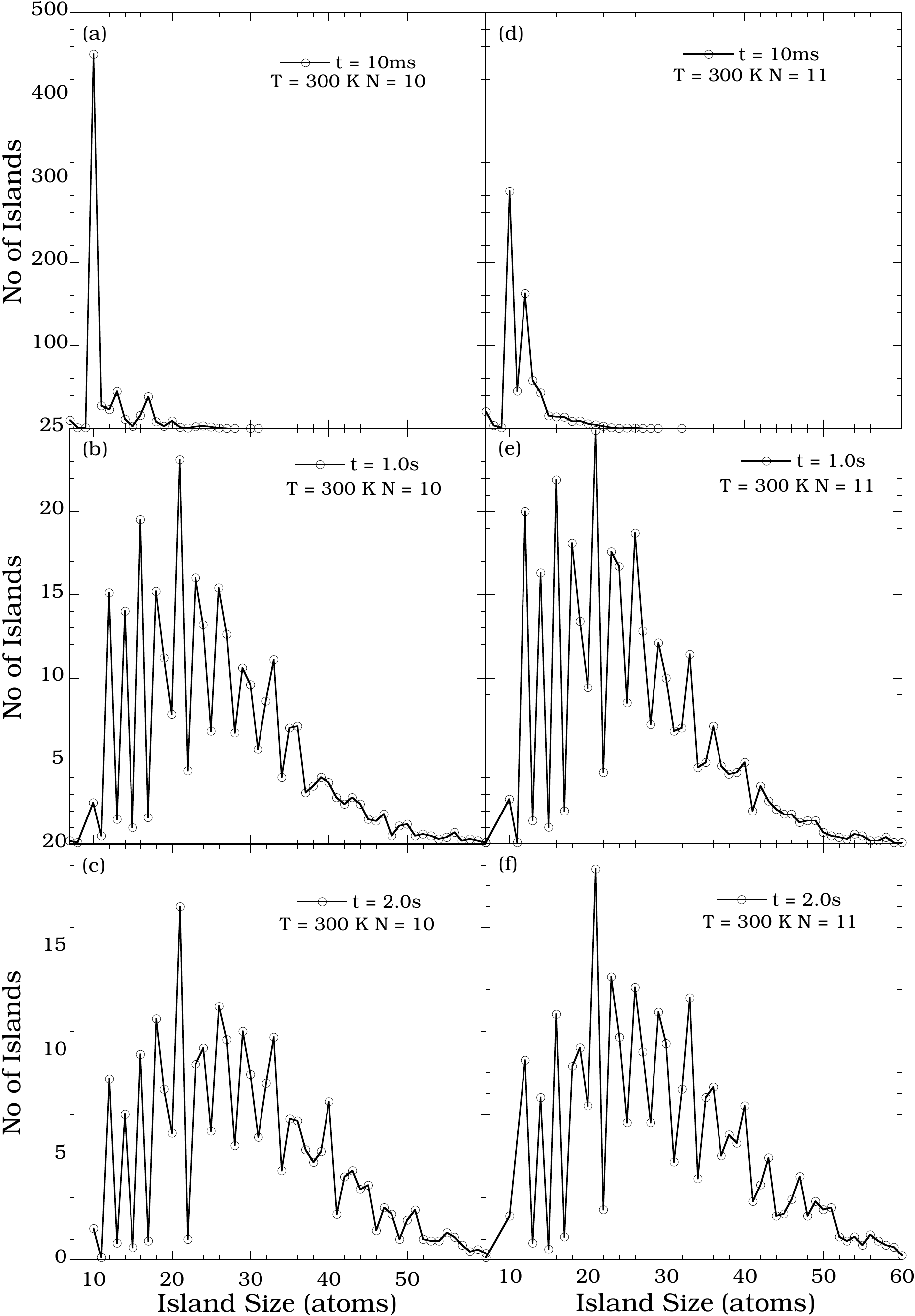} 
\caption{\label{isd_300k_delta}{ISDs at T = 300 K when the initial ISD is a delta function in which all 752 islands are of size 10 or size 11, respectively:  at (a)/(d) 10 ms; (b)/(e) 1.0 s; (c)/(f) 2.0 s}}
\end{figure}

 \begin{figure}
 \includegraphics [width=8.5cm]{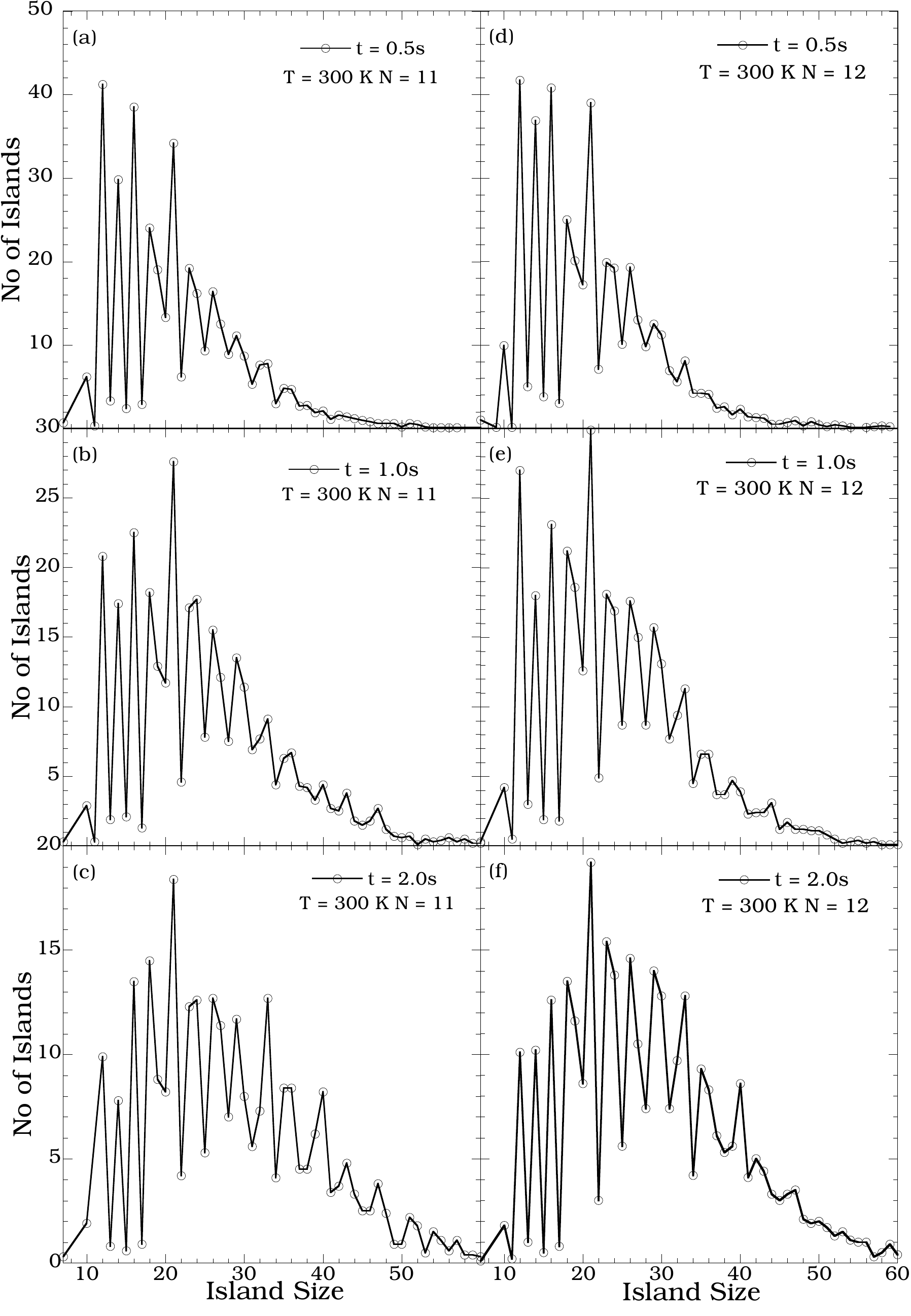} 
\caption{\label{isd_300k_gaussian}{ISDs at T = 300 K when the initial ISD is a Gaussian function in which average island size is 10 atoms or 11 atoms, respectively:  at (a)/(d) 10 ms; (b)/(e) 1.0 s; (c)/(f) 2.0 s}}
\end{figure}

We found that during early stages when island sizes are small coarsening proceeds as sequence of selected island sizes, \cite{iss_short}  as is clearly evident in Figs.~\ref{isd_300k_delta} \& \ref{isd_300k_gaussian}. 
which  show ISDs at various times when the starting ISDs are delta and Gaussian respectively. We carried out these simulations up to 3.0 s. It can seen that as the coarsening proceeds there is a dramatic change in the ISD from a sharp delta or a smooth Gaussian to a non-smooth distribution, with peaks and valleys.
Table~\ref{table1} summarizes island sizes up to  $35$ atoms after $1.0$ s of coarsening according to whether they constitute a peak, valley or neither within ISD. It can be seen in Fig.~\ref{isd_300k_delta} \& \ref{isd_300k_gaussian} that at some island sizes ($19$, $27$ and $30$ atoms) there is neither a peak not a valley within the ISD. 
At either size $23$ or size $24$ can there be a peak, though for the most part, the peak occurs at size $23$, while at size $24$ there is neither a peak not a valley.
Note that at much later times all island coarsening exhibits Ostwald ripening, resulting in a single large island:  the total energy of the system will decreases as more bonds are formed until it saturates when one large island is formed.
These characteristics of early- and late-stage coarsening are independent not only of whether the initial ISD is Gaussian or delta but also of wether it is random. We confirmed the latter by carrying out coarsening simulations with an initial ISD created by depositing Ag atoms on Ag (111) surface at very low temperature ($135$ K) with a slightly higher monomer diffusion barrier to increase the number of islands and to keep the average island size smaller. Under these conditions islands are fractal and the ISD is random.
The ISD exhibits the same characteristic change even when the shapes of islands are altered in the initial configuration: all islands with a kinetically stable shape or a low-energy or an irregular (e.g fractal) shape.  We also found that, although the strength of selectivity depends on temperature, however, there is no change with temperature in the sizes of islands that constitutes peaks and valleys in the ISD.
In sum (as out earlier study showed)  island-size selectivity is independent of all parameters of initial ISD, including temperature, island shape(s), and whether the distribution is random, delta, or Gaussian.

 \begin{table}
\caption{\label{table1} List of islands size for which the ISD is a peak, a valley or neither after $3.0$ s coarsening.}
 \begin{tabular}{ c | c c c c c c c c c c}
\hline
\hline
Feature & & & &Sizes& of & Islands\\
\hline
Valley & 11~ & 13 ~& 15 ~& 17~ & 20 ~& 22 ~& 25 ~& 28 ~& 31 ~& 34\\ 
Peak & 12 ~& 14 ~& 16~ & 18 ~& 21 ~& 23(24)~ & 26~ & 29~ & 33 ~& 35\\
 neither& & & & 19 ~& & & 27~ & 30~ & 32\\
 \hline
 \hline
\end{tabular}
\end{table}


\begin{figure}
\includegraphics [width=8.5cm]{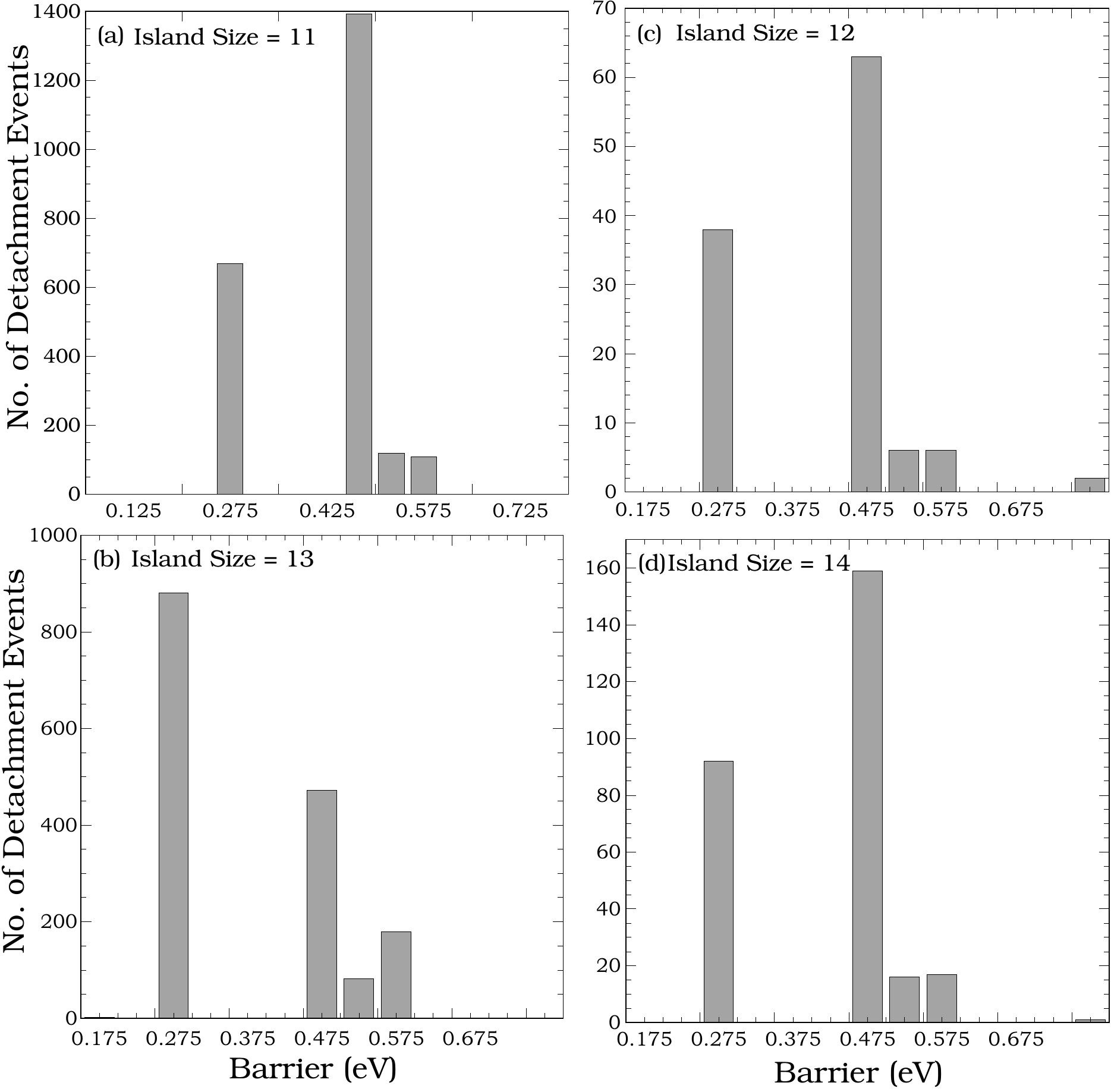} 
\caption{\label{barrier_bin}{Histogram of detachment events selected for islands of size (a) $11$ atoms (b) $12$ atoms (c) $13$ atoms and (d) 14 atoms during of $3$ seconds of coarsening. Histogram interval is $0.05$ eV.}}
\end{figure}

\begin{figure}
\includegraphics [width=8.5cm]{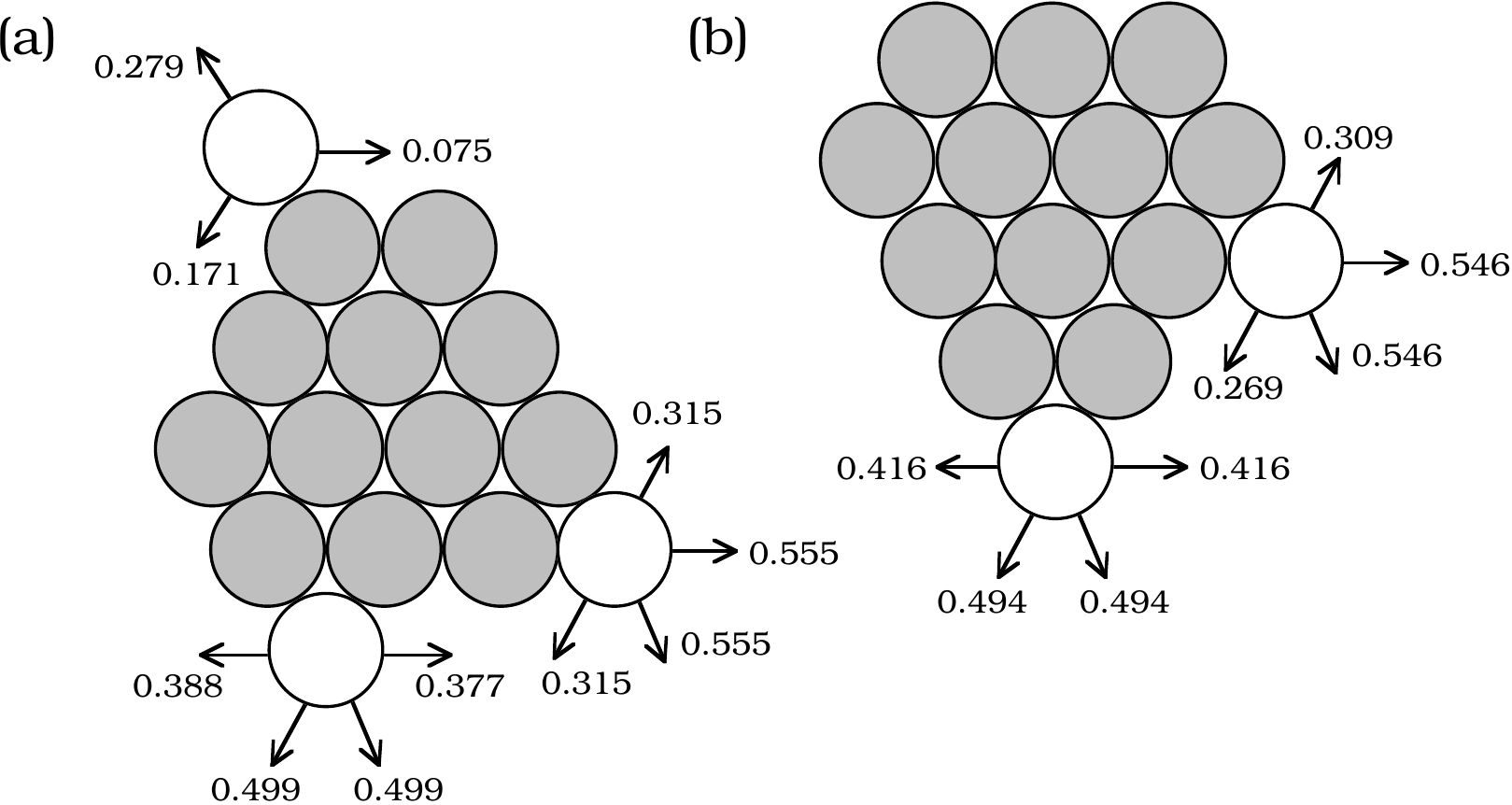} 
\caption{\label{detach_processes}{Activation barriers (in eV) for the most frequent detachment processes and for edge diffusion processes.}}
\end{figure}

It is known \cite{morgenstern1, OstwaldAg,pslkmc}  that for two-dimensional Ag late-stage coarsening on Ag(111) surface is due to evaporation and condensation mediated by monomer diffusion between islands. 
To elucidate the factors that determine the pattern of island-size selectivity described above during early-stage coarsening, we examined the energetics of detachment processes on the basis of island size up through island size $21$, starting from island size $8$.
(For islands whose sizes are smaller than $8$ atoms, the energy barrier for the most frequent concerted diffusion processes are quite small [$0.1-0.3$ eV], causing these islands to diffuse and coalesce with others.)  
Fig.~\ref{barrier_bin} shows histograms of energy barriers for detachment events selected during $3.0$ s of coarsening for island sizes $11$ through $14$ when the initial ISD was a Gaussian. 
Note that one can discover a one-to-one correspondence between the histograms in Fig.~\ref{barrier_bin}  and the actual calculated barriers for the most frequent edge-atom detachment processes illustrated in Fig.~\ref{detach_processes}.
That is: In Fig.~\ref{barrier_bin}, $0.275$ eV corresponds to the activation barrier for detachment of a corner atom, $0.475$ eV corresponds to that for detachment of an edge atom from a B-type step edge; and barriers $0.525$ eV and $0.575$ eV correspond to those for detachment of an edge-atom from an A-type step edge. 
From Fig.~\ref{detach_processes} it can also be seen that the difference between an edge diffusion barrier and an edge-atom detachment barrier is quite small on a B-type step edge, making  it a most frequent type of detachment process. 
Recall (from Figs. 2 and 3) that the populations of islands of sizes 11 and 13 atoms constitute valleys while that of islands of sizes 12 and 14 atoms constitute peaks in the ISD.   
Now compare the height of the histogram in Fig.~\ref{barrier_bin}(a) with that in Fig.~\ref{barrier_bin}(c) and that in Fig.~\ref{barrier_bin}(b) with that in Fig.~\ref{barrier_bin}(c), we see that the number of events of edge-atom detachment for island sizes whose populations constitute valleys in the ISD is much higher than that for island sizes whose populations constitute  peaks in the ISD.
The reason is that because the energy barrier for an atom with at least $3$ nearest neighbor atoms to detach is greater than $0.7$eV, such atoms rarely detach to create monomers at room temperature.
These arguments based on system energetics confirm findings in our KMC simulations that island of sizes whose populations are peaks in the ISD have no edge atoms: instead, all their atoms have at least $3$ nearest neighbor atoms, making them kinetically stable islands.  We also found that artificially  increasing the energy barrier for the most frequent detachment processes delays the onset of island selectivity to later times.
Accordingly, we conclude that island-size selection is primarily due to adatom-detachment and attachment processes at island boundaries owing to the relative ease with which edge atoms can detach in comparison with the relative difficulty for the detachment of atoms with at least $3$ nearest neighbor atoms.

\subsection{Shape Analysis}\label{shape}
In order to gain further insight in to island-size selectivity we looked at shapes of islands as a function of their size.  Shapes an island can assume during coarsening are constrained by the fact that detachment events are predominately edge-atom detachments and rearrangement of atoms in an island rarely happens due to high detachment barrier for atoms with at least $3$ nearest neighbors. We use the term "compact shape" to specify a shape with a closed-shell structure, i.e., one with no edge-atoms or kinks. As already mentioned, islands of sizes smaller than $8$ atoms are hardly present  since they quickly diffuse and coalesce with other islands. Therefore we follow the shapes starting from islands of size $10$ atoms  which is the first stable size after islands of size $8$ atoms, all the way up to islands of size $21$ atoms. 
  In order to uniquely identify the shape of the island on-the-fly, we used following $3$ criteria, ($1$) number of nearest neighbors, ($2$) maximum distance of any atom from the center of mass of the island  and ($3$) maximum distance between any two atoms in the island. Most often criteria $1$ \& $2$ are sufficient to uniquely identify the shape of an island.

\begin{figure}
\includegraphics [width=6.5cm]{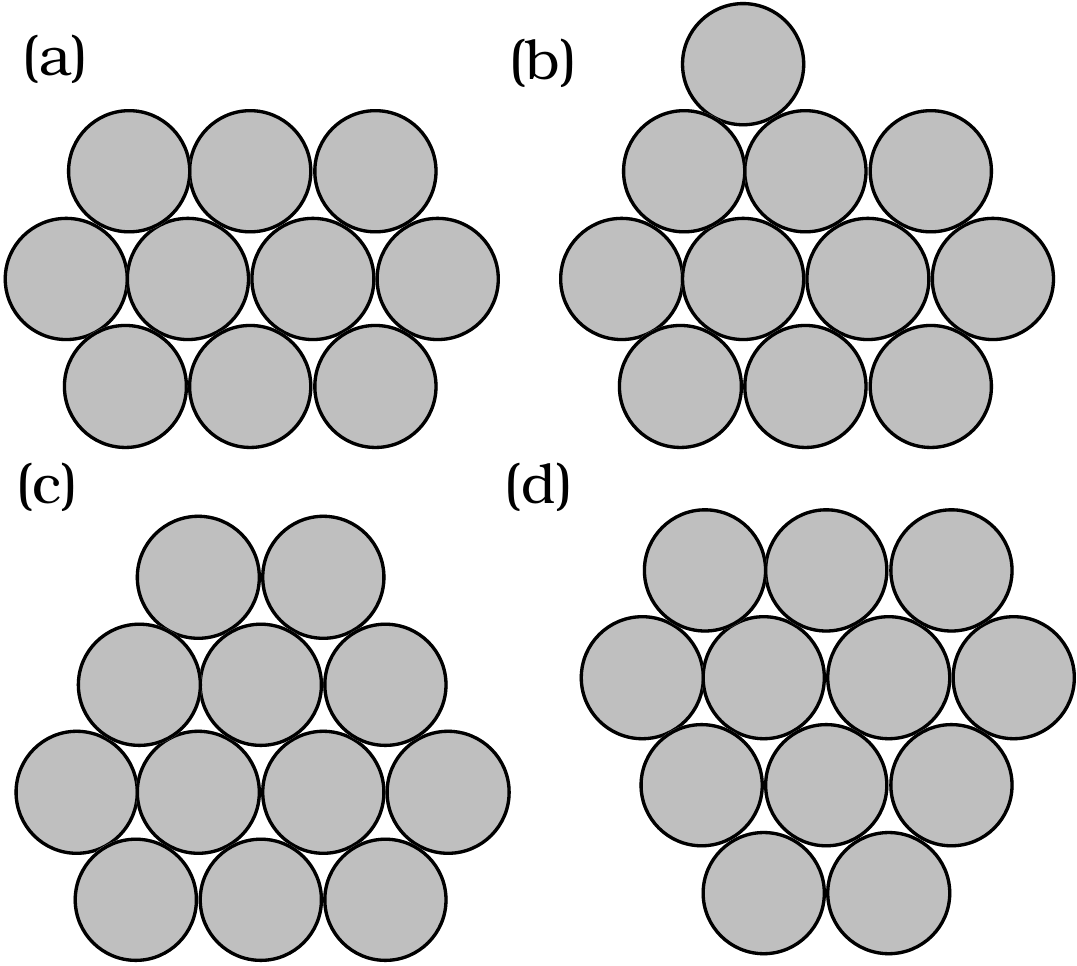}
\caption{\label{Is10-11-12}{Most frequent shapes for islands of size (a) 10 atoms, (b) 11 atoms, (c) 12 atoms,  and (d) 12 atoms.  Note that in (c) the B-type step edge is longer than the A-type, while in (d) the opposite is the case.}}
\end{figure}

Fig.~\ref{Is10-11-12} (a) shows the most frequent shape observed for islands of size $10$ atoms during coarsening. There are two other possible orientations for this shape which can be obtained by the rotating  shape shown in Fig.~\ref{Is10-11-12}(a) by $120^{0}$ or $240^{0}$ either clockwise or anti-clockwise about the center of mass of the island. For brevity we will show  only one of the possible orientations. 
We note that barrier for a monomer to attach to an A-type step edge ($0.04$ eV) is smaller than the barrier for it to attach to a B-type step-edge ($0.06$ eV). Also it can be seen from Fig.~\ref{detach_processes} that barrier for edge atom to detach from a B-type step-edge is slightly smaller than an A-type step-edge. Since it is easier for monomer to attach and harder to detach for an A-type step-edge, the most frequent shape observed for islands of size $11$ atoms is as shown in Fig.~\ref{Is10-11-12}(b), with an edge-atom on A-type of step edge.
In contrast, $11$ atom island with an edge-atom on B-type step edge can also form during coarsening, but does not survive very long compared to the shape shown in  Fig.~\ref{Is10-11-12}(b). It will quickly changes to island of size $10$ atoms through detachment or rarely to a $12$ atom island due to attachment of a monomer.
Similarly, fig~\ref{Is10-11-12}(c) shows the most frequent shape observed for islands of size $12$ atoms, forms when a monomer attaches to the kink on the most frequent shape of island of size $11$ atoms, resulting in a longer B-type of step edge. Fig.~\ref{Is10-11-12}(d) shows the next most frequent shape for islands of size $12$ atoms, with longer A-type step edges which is formed when a monomer attaches to the kink on B-type step-edge on the less frequent shape of $11$ atom island. Since the two most frequent shapes for islands of size $12$ atoms are compact and survive much longer, their population constitutes a peak in the ISD

\begin{figure}
\includegraphics [width=6.5cm]{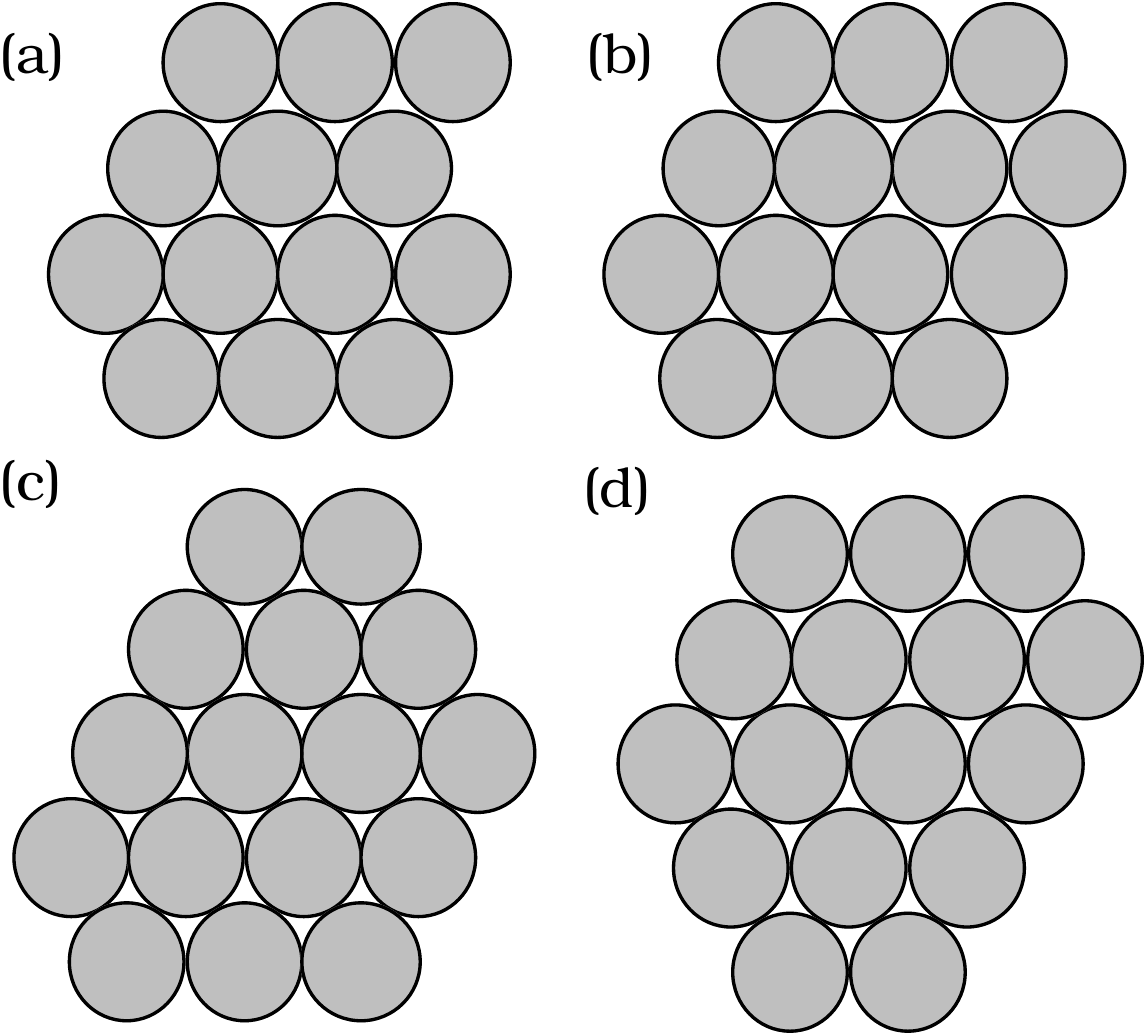} 
\caption{\label{Is14_shapes}{Shapes of Islands of sizes (a) $13$ atoms (b) $14$ atoms (c) $16$ atoms (longer B-type step-edge) and (d) $16$ atoms (longer A-type step-edge)}}
\end{figure}

Figure.~\ref{Is14_shapes}(a) shows the most frequent shape observed for islands of size $13$  atoms during coarsening. It can be seen that its shape is a combination of $12$ atom island with a longer B-type step and one extra edge atom. This edge atom can either easily detach to form island of size $12$  atoms or a monomer can attach to form island of size $14$ atoms, so that the population of islands of size $13$ atoms constitute a valley in the ISD.
Fig.~\ref{Is14_shapes}(b) shows most frequent  shape for island of size $14$ atoms. This shape and its other two orientations (rotate by $120^{0}$ and $240^{0}$) can also be formed by attaching $2$ atoms to the step edges to either of two shapes shown in Figs.~\ref{Is10-11-12}(c) \& (d) for islands of size $12$ atoms  resulting in the same shape for islands of size $14$ atoms,  with $2$ longer A-type and B-type step edges. 
Figs.~\ref{Is14_shapes} (c) \& (d) shows the two most frequent shapes for islands of size $16$ atoms, one with a longer B-type step edge and the other with longer a A-type step-edge, respectively.  
Shape of $16$ atom islands shown in Figs.~\ref{Is14_shapes}(c) \& (d) are formed when two atoms attach to A-type and B-type step-edges of a $14$ atom island shown in Fig.~\ref{Is14_shapes}(b), respectively.  As mentioned earlier, since it is easier for a monomer to attach to an A-type step-edge and harder to detach for it compared to a B-type step-edge, shape of $16$ atom island shown in Fig.~\ref{Is14_shapes}(c) is the most frequent observed shape.
Since the most frequent shapes for island of size $16$ atoms are compact, their populations constitutes a peak in the ISD and any monomer that attaches to the step-edge easily detaches, with the consequence that the population of islands of size $17$ atoms constitutes a valley in the ISD.

\begin{figure}
\includegraphics [width=6.5cm]{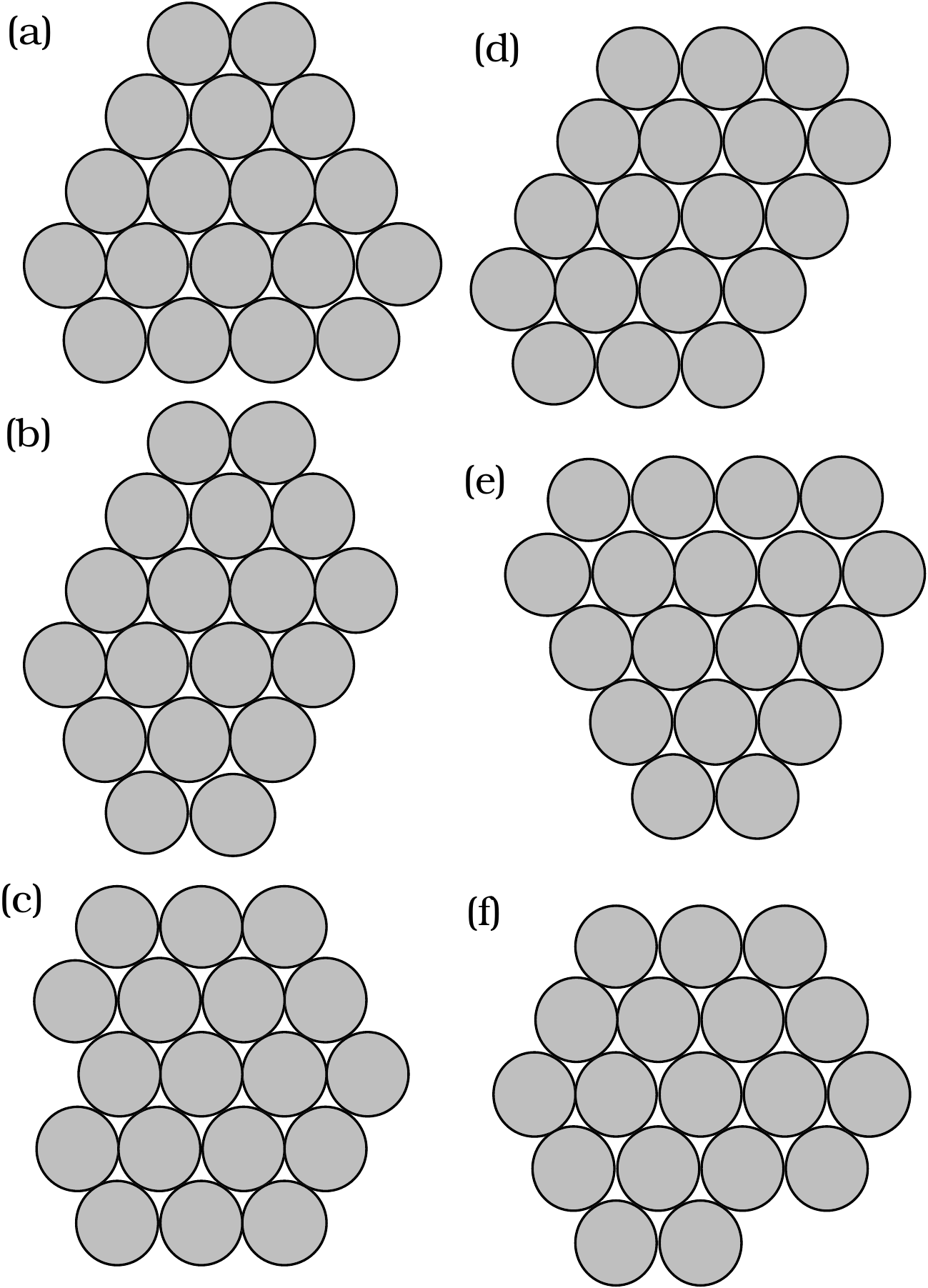} 
\caption{\label{Is18_shapes}{Most frequent shapes observed for islands of size  $18$ atoms.}}
\end{figure}

Fig.~\ref{Is18_shapes} shows all the most frequent shapes observed for islands of size $18$ atoms during coarsening. These shapes are formed when $2$ atoms attach to one of the step-edges of islands of size $16$ atoms shown in Figs~\ref{Is14_shapes}(c) \& (d).  The most frequent  shapes for islands of  size $18$ atoms are the compact and therefore their population constitutes a peak in the ISD.
Since islands of size $18$ atoms are compact, one expects population of islands of size $19$ atoms constitute to be a valley in the ISD. But islands of size $18$ atoms, though less frequently, also form non-compact shapes shown in Figs.~\ref{Is18_shapes}(c) \& (f) which  quickly absorb a monomer to form a kinetically stable hexagonal shapes $19$ atom island. Since these islands are extremely stable once formed, they are most frequently observed type of $19$ atom islands.
Accordingly, as can be seen from Figs.~\ref{isd_300k_delta} \& \ref{isd_300k_gaussian} that the population of islands of size $19$ atoms constitutes neither a peak nor a valley in the ISD instead of being a valley in the ISD.
We note that Fig.~\ref{Is18_shapes}(a) is the most frequent shape for islands of size $18$ atoms while shapes in Fig.~\ref{Is18_shapes}(b) \&(d) are less frequently observed compact shapes.

\begin{figure}
\includegraphics [width=6.0cm]{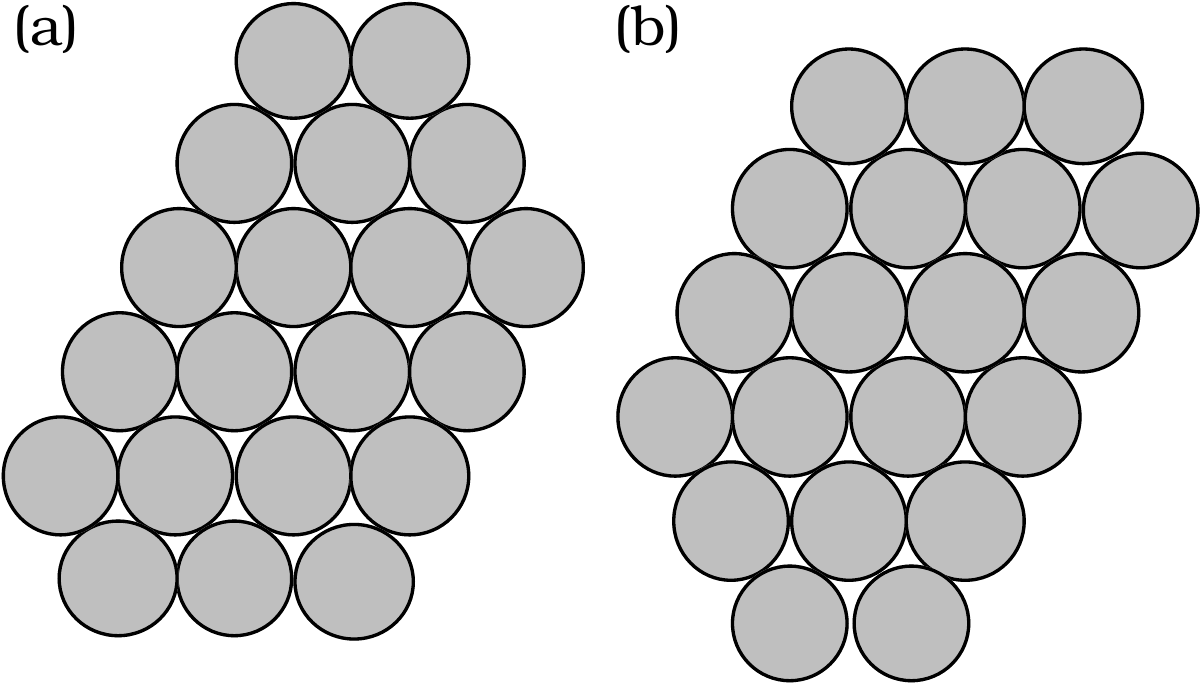} 
\caption{\label{Is20_shapes}{Most frequent shapes of islands of size $20$ with longer (a) B-type (b) A-type step edges}}
\end{figure}

Figs.~\ref{Is20_shapes}(a) \& (b) show compact shapes observed for islands of size $20$ atoms during coarsening. These shapes are formed when two atoms attach to the shorter step-edge (3-atoms wide) of those islands of size $18$ shown in Figs~\ref{Is18_shapes} (b) \& (d). 
Other shapes that form most often for islands of size $20$ atoms during coarsening are not compact: they have either an edge-atom or a kink. 
These non-compact shapes quickly change to islands of either size $18$ or $19$ atoms due to detachment or size $21$ atoms due to attachment of a monomer. 
Fig.~\ref{Is21_shapes} shows the most frequent shapes for islands of size $21$ atoms observed during coarsening. It can be easily checked that these shapes can be obtained by attaching $3$ atoms to longer edges in Fig.~\ref{Is18_shapes} (a), (b), (d) \&(e) and $2$ atoms to any step-edge of hexagonal shaped $19$ atom island. In particular, all shapes shown in Fig.~\ref{Is18_shapes} lead to shapes shown in Fig.~\ref{Is21_shapes} for islands of size $21$ atoms. As can be seen in Figs.~\ref{isd_300k_delta} \& \ref{isd_300k_gaussian}, this leads to a very sharp population peak at islands of size $21$ atoms in the ISD.

\begin{figure}
\includegraphics [width=4.0cm]{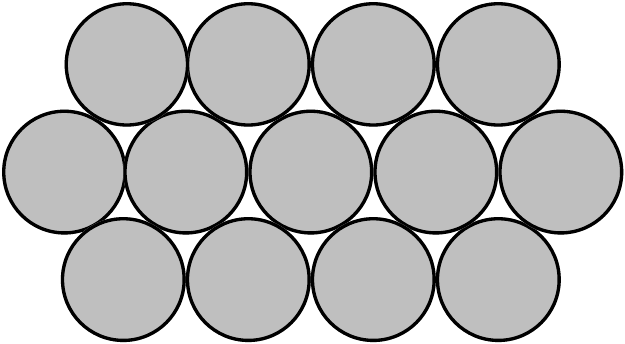} 
\caption{\label{Is13stable}{Kinetically stable shape for island size $13$}}
\end{figure}

Due to high detachment barrier  for atoms  with at  least 3 nearest  neighbors, they rarely detach to create monomers and atoms in an island do not rearrange  except through  attachment and detachment  of edge atoms.   Therefore,  whether  an island takes  a particular  shape or not  depends on the  shapes of smaller islands formed during  coarsening.  Moreover, certain non-selected  island sizes, even though  a kinetically stable  shape might exist for such island size, appear only rarely, since except through detachment  atoms never rearrange  themselves into a kinetically stable shape.  For example, for island of size $13$ atoms Fig. 7(a) shows the most frequent shape observed during coarsening while Fig. 11 shows kinetically stable shape possible. For islands of size $20$ atoms due to the presence of compact island shapes it can be seen that  its population  density is non-zero even though it constitutes a valley in the ISD. It can also be seen in Figs.~\ref{isd_300k_delta} \& \ref{isd_300k_gaussian}, islands of sizes $11$, $13$, $15$ and $17$ atoms, which never form a compact  shape during  coarsening, usually have zero population  density.  For islands of size $11$, $15$ and $17$ atoms, compact  shapes are a geometric impossibility.  Accordingly we conclude that  island  selection is primarily  due to edge-atom detachment  and attachment  processes at island boundaries  owing to the relative ease with which atoms can detach in comparison with the relative difficulty for the detachment of atoms with at least $3$ nearest neighbors.


\begin{figure}
\includegraphics [width=7.0cm]{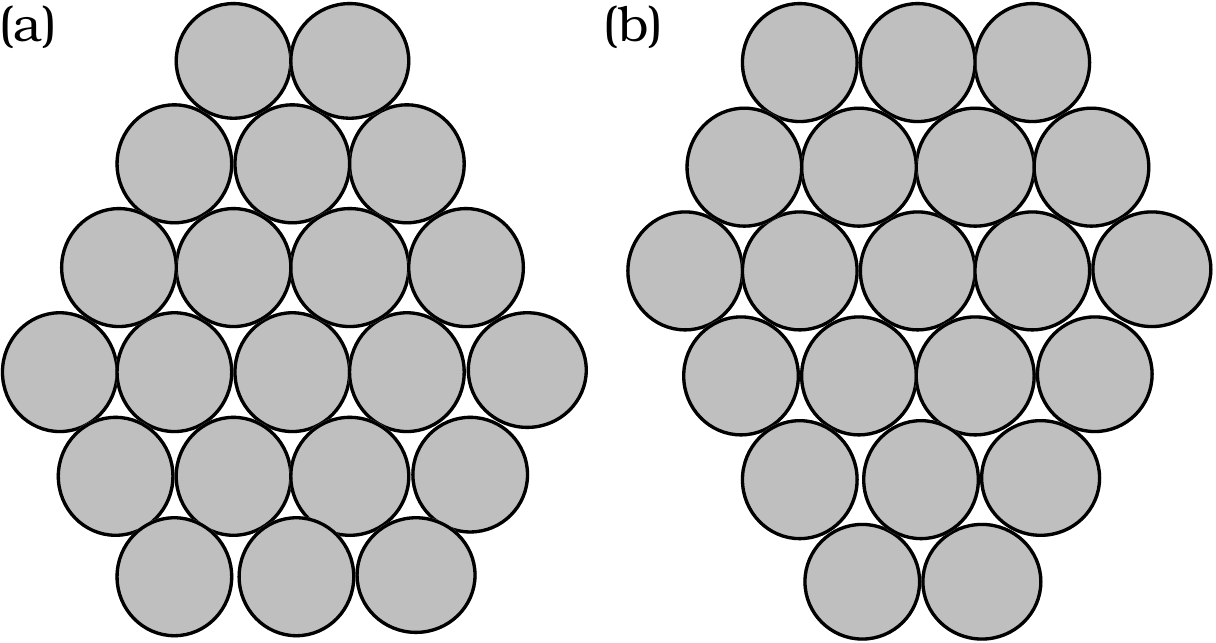} 
\caption{\label{Is21_shapes}{Most frequent shapes of islands of size $20$ with longer (a) B-type (b) A-type step edges }}
\end{figure}

\section{Conclusions} \label{conclusions}

We find that,  during  early stages,  that  is when island sizes are small, two-dimensional Ag island  coarsening  on Ag(111) surface  proceeds  as a sequence of selected  island  sizes resulting  in peaks  and  valleys in the ISD. Densities of islands of selected  sizes decay at slower rates  because of the  formation  of kinetically  stable  island shapes, while densities  of non-selected  sizes (valleys in the ISD) decay more rapidly owing to non-formation of selected island shapes. In order to understand the reasons we have studied in detail the island shapes both for selected and non-selected island sizes observed during coarsening. A kinetically stable shape has a closed shell structure with all periphery atoms having at least three nearest neighbor bonds and with no kinks, thus making the detachment of a periphery atom a rare process, consequently their densities decays at a slower pace. In contrast densities of non-selected sizes decay more rapidly owing to a higher frequency of edge-atom attachment/detachment processes  which are either single or doubly bonded atoms and easier to detach resulting the creation of monomers.

Although  kinetically unstable shapes of a selected island size are formed during coarsening, their densities are negligible compared to the density of kinetically stable shapes. Furthermore it is also possible for certain non-selected island sizes to have a kinetically stable shape, but they rarely form one while for the rest of non-selected sizes kinetically stable shape is geometrically not possible and therefore always have population densities close to zero.  
In case of kinetically stable islands, since they do not have any kink in their shapes, it is difficult for a periphery atom to detach even to form an edge diffusing atom, these islands once formed do not rearrange into other shapes. 
But if there is a kink in the shape of an island, a kink atom whose activation barrier to detach along the edge to form an edge-atom is slightly larger than the barrier for an edge-atom to detach to form a monomer, may result in the formation of an edge-atom which can either detach to form monomer or the kink may disappear by the attachment of an edge-atom created due to the attachment of a monomer.   
As the coarsening proceeds island sizes gets bigger and number of shapes an island can assume become larger and each island size can have multiple kinetically stable shapes. In our coarsening simulations we have found that island selectivity is clearly visible until island of $30-40$ atoms while for the larger island sizes their population densities are so small that selectivity cannot be observed.

Coarsening results presented in this article were started with ISDs created manually with Gaussian distribution. We have still observed same island-size selectivity when the initial ISD was created by deposition at a very low temperature and islands formed have fractal shapes.
We have observed that the island-size selectivity was independent of parameters of initial ISD like average island size, island shapes and the type of distribution showing that it is a characteristic of early stage of Ag island coarsening on Ag(111).  In addition we also found that  though  island-size  selectivity was independent of temperature how strong  it  appears  depends  on  temperature, strongest  between  $250 - 270$K. Island-size selectivity was observable until $300$K  beyond this temperature island densities decays so fast that selectivity was barely observable. In conclusion island-selectivity in case of Ag island coarsening on Ag(111) surface depends on attachment/detachment processes and it would in future be interesting to do a detailed study on how island-size selectivity depends on various types of processes and their activation barriers.



\begin{acknowledgments}

This work was supported by NSF grant ITR-0840389. We would also like to acknowledge computational resources provided by University of Central Florida. We thank Lyman Baker for critical reading of the manuscript.

\end{acknowledgments}

\bibliography{references}

 \end{document}